\documentclass{an}
\usepackage{graphicx}
\usepackage{times}
\Volume{01}              % issue in current volume (01...06)
\Year{2002}              % format: {YYYY}
\Month{10}               % format: {M} or {MM} (number), month of
\Pagespan{223}{227}      % format: {start page}{last page}

\begin{document}
\lhead[\thepage]{A. Comastri: Unconventional AGN}
\rhead[Astron. Nachr./AN~{\bf XXX} (200X) X]{\thepage}
\headnote{Astron. Nachr./AN {\bf 32X} (200X) X, XXX--XXX}

\title{Unconventional AGN in hard X-ray surveys} 

\author{A. Comastri\inst{1}
\and  M. Brusa\inst{2,1}
\and M. Mignoli \inst{1}
\and the {\tt HELLAS2XMM} team \fnmsep\thanks{A. Baldi, P. Ciliegi, 
F. Cocchia, F. Fiore, F. La Franca, R. Maiolino, G. Matt, S. Molendi,  
G.C. Perola, C. Vignali}}
\institute{
INAF - Osservatorio Astronomico di Bologna, via Ranzani 1, 40127 Bologna
Italy 
\and 
Dipartimento di Astronomia, Universit\`a di Bologna, via Ranzani 1, 
40127 Bologna, Italy}

\date{Received {\it date will be inserted by the editor}; 
accepted {\it date will be inserted by the editor}} 

\abstract{Extensive programs of follow--up observations of hard 
X--ray selected sources have unambiguosly revealed that the 
sources of the X--ray background
are characterized by an extremely large dispersion in their 
optical magnitude and spectroscopic classification.
Here we present the results of an attempt to understand 
the nature of the observed variety using a simple prescription for 
their optical to X--ray energy distribution.
}

\correspondence{comastri@anastasia.bo.astro.it}

\maketitle

\section{Introduction}

Thanks to the capabilities of the detectors onboard {\it Chandra} and 
XMM--{\it Newton} the X-ray sky is now probed down to flux 
limits where the 
bulk of the hard X-ray background is resolved into single sources.
The accuracy in the positioning of hard (2--10 keV) 
X--ray sources is more than an order of magnitude 
better than that achieved by ASCA and BeppoSAX 
in the same energy range.
One of the most interesting consequences of this enormous improvement
is the possibility 
to identify a relatively large number
of hard X--ray sources characterized by broad band properties
which are significantly different from those of {\tt conventional} 
AGN selected in the optical and soft X--ray bands.
Although there are compelling theoretical and observational evidences  
which suggest that the large majority of the 
hard X--ray sources are obscured AGN, the origin of such a broad variety 
in their multiwavelength properties is still far to be understood.  

\section{The $f_X/f_{opt}$ diagnostic}

It is well known that various classes of X--ray emitters are
characterized by different values of their 
X--ray to optical flux ratio (see fig. 1 in Maccacaro et al. 1988).
For a given X--ray energy range and the R band filter the 
following relation holds:  $log f_X/f_{opt} = log f_X + 5.5 + R/2.5$.
The large majority of spectroscopically identified AGN 
in both ROSAT  (e.g. Hasinger et al. 1998) and ASCA (Akiyama et al. 2000) 
surveys fall within $-1 < log(f_X/f_{opt}) < $ 1.
Extensive optical follow--up observations of hard X--ray sources 
discovered by deep and medium deep {\it Chandra} and XMM--{\it Newton}
surveys confirm this trend to fainter X--ray fluxes and, at the same 
time, show evidence of a relatively large number of sources which deviate 
from log$(f_X/f_{opt}) = 0\pm1$ (Fig.~1).
For the purposes of the present paper it is convenient 
to divide the ``outliers'' in two groups.
The first includes sources that are X--ray weak for their 
R band magnitudes (log$f_X/f_{opt} \simeq$ --1); the second, 
sources that are optically faint 
(sometimes below the limits of deep optical images) and relatively
X--ray bright  (log$f_X/f_{opt} > $ 1).
In the following we refer to both classes of 
sources as {\tt unconventional} AGN. 

\begin{figure*}
\resizebox{\hsize}{!}
{\includegraphics[width=17cm,height=18cm]{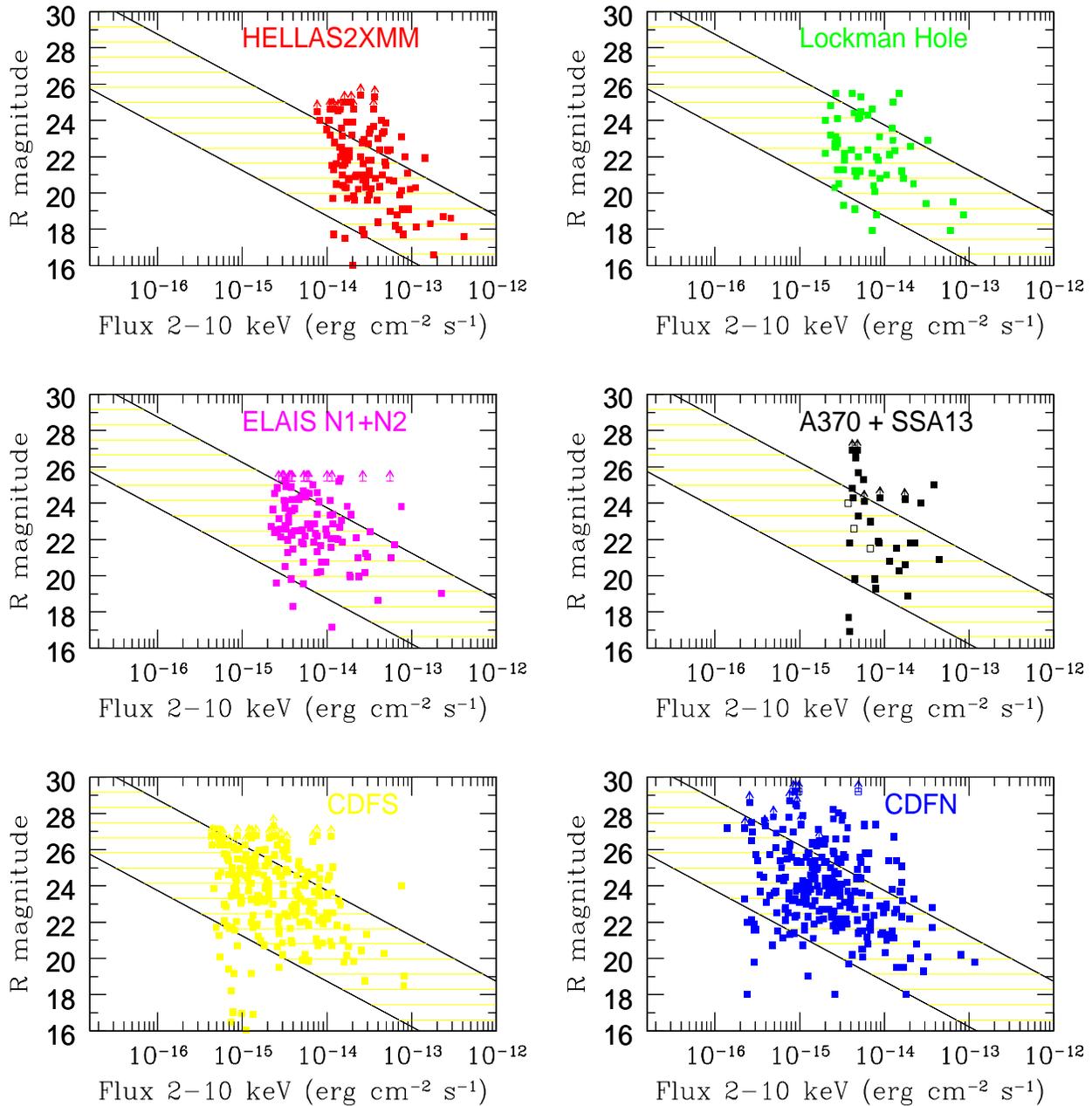}}
\caption{\small The 2--10 keV flux versus the R band magnitude for six
different surveys as labeled. From top left 
the {\rm HELLAS2XMM} survey (Baldi et al. 2002), The XMM survey in the 
Lockman Hole (Mainieri et al. 2002); the {\rm ELAIS} deep {\it Chandra} 
survey (Manners et al. 2002), the combination of a few medium deep 
{\it Chandra} surveys (Barger et al. 2001), The {\it Chandra} Deep Field 
South (Giacconi et al. 2002) and the {\it Chandra} Deep Field 
North (Brandt et al. 2001). The upper (lower) solid line corresponds to 
log$f_X/f_{opt}$=1 (--1). The shaded area represents the region occupied 
by {\tt conventional} AGN (e.g. quasars, Seyferts, emission line galaxies).}
\label{label1}
\end{figure*}

The identification breakdown of the sources in the first group 
is a mixed bag including emission line galaxies
and apparently normal galaxies (see Alexander et al. this volume).
A sizeable fraction of the latters, named {\tt XBONG} 
(X--ray Bright Optically Normal Galaxies; Comastri et al. 2002b),
are particulary intriguing being characterized by 
an absorption dominated optical spectrum and AGN--like hard 
X--ray luminosities ($L_{2-10} \simeq 10^{42-43}$ erg s$^{-1}$). 
They are found at moderately low redshift, ($z<$ 1; 
Hornschemeier et al. 2001, Barger et al. 2002). The average value
of their log$f_X/f_{opt}$ distribution is around --1 with a 
large dispersion (Fig.~3).
An attempt to deeply investigate their nature through a multiwavelength 
approach suggests that the putative AGN responsible
for the hard X--ray emission is completely hidden at longer wavelengths 
(Comastri et al. 2002a).

The sources characterized by high values of $f_X/f_{opt}$ are even 
less understood. The spectroscopic identification of these objects
is already challenging the capabilities of 8--10 m optical telescopes 
calling for the next generation of ground--based facilities or for
alternative techniques such as multicolour optical photometry 
and, for the X--ray brightest sources, on the search for redshifted iron 
K$\alpha$ lines.
In this respect it is important to note that wide area, shallow, hard X--ray
surveys, designed to sample relatively bright X--ray and optical fluxes (Fig.~1),
are best suited to investigate the nature of sources with 
high X--ray to optical ratios. 
Indeed a few examples of objects with log$f_X/f_{opt} >  1$ 
in the {\tt HELLAS2XMM} (F. Fiore et al., in preparation) 
and {\it Chandra} surveys (Fabian et al. 2000) turned out 
to be high redshift, highly obscured AGN.  

\begin{figure}
\resizebox{\hsize}{!}
{\includegraphics[]{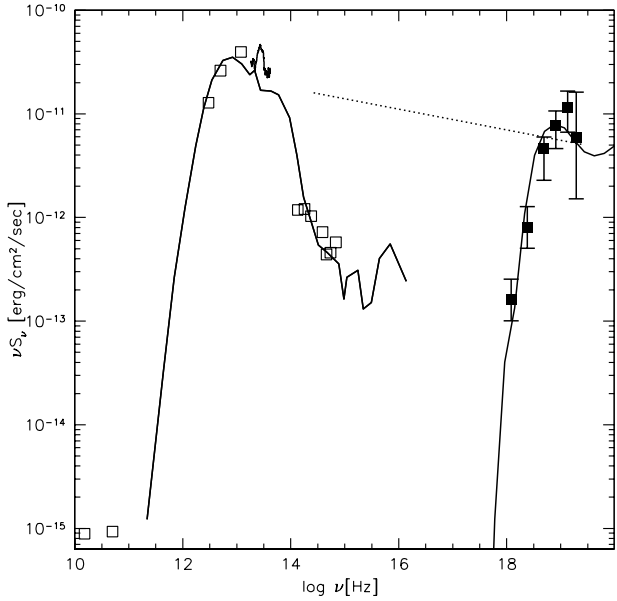}}
\caption{\small The broad band spectral energy distribution of IRAS 09104+4109 
(adapted from Franceschini et al. 2000).}
\label{label2}
\end{figure} 

Although obscured accretion seems to provide 
the most likely explanation for the optical and X--ray flux distribution  
of {\rm unconventional} AGN, alternative possibilities are viable.
At very faint optical fluxes (log $f_X/f_{opt} > 1$) the observed properties 
would be consistent with both high redshift (possibly even $z > 6$) quasars 
and high--z ($> 1$) cluster of galaxies. 
The X--ray emission in optically bright sources ($f_X/f_{opt} \simeq -1$) 
could be ascribed, at least in part, to star--formation 
related processes (see Hornschemeier 2002) or, for those objects detected 
in the radio band, to a BL Lac object with rather extreme properties
              (M. Brusa et al. in preparation).

\section{A toy model for unconventional AGN}

In the following we try to further investigate the nature 
of {\tt unconventional} AGN assuming that they are powered 
by heavily obscured accretion.
In order to properly address this issue it 
is important to point out that as long as the absorption 
column density 
does not exceed values of the order of ``a few'' $\times$ 10$^{24}$ cm$^{-2}$
(mildly Compton--thick) the 
high energy spectrum recovers at $E > $10--20 keV  
(see fig.~2); for higher values of the intrinsic absorption
(heavily Compton--thick), 
Compton down--scattering strongly suppress the nuclear radiation 
at all energies and only reflected light with a quite flat 
slope can be seen (Matt et al. 1999). 
A strong iron line (EW $>$ 1 keV) is expected in both the cases.  
At longer wavelengths the nuclear radiation is completely blocked 
by dust and the optical infrared spectrum is dominated by the 
host galaxy starlight.
There are several examples of both mildly and heavily 
Compton thick sources in the local Universe discovered 
thanks to BeppoSAX.
The prototype of the former is NGC 6240 at $z=0.0245$ (Vignati et al. 1999)
while for the latter an handful of objects are reported 
by Maiolino et al. (1998). 
Whether Compton thick AGN are common at high redshift is still debated
(Fabian et al 2002) though the ultraluminous infrared 
quasars IRAS 09104$+$4109 at $z=0.442$ (Franceschini et al. 2000, 
Iwasawa et al. 2001) and F15307 at $z=0.92$ (Ogasaka et al. 1997) are 
good examples of mildly and heavily Compton thick AGN, respectively.

It is quite obvious to note that moving the SED of 
a mildly Compton thick object (Figure 2) 
to progressively higher redshifts the K--corrections in the optical and 
X--ray band work in the opposite direction.
The shape of the hard X--ray spectrum 
is responsible of the strong positive K--correction which   
``boosts'' the X--ray flux and favours the detection of high redshift sources.
The same effect has been already extensively used 
to detect high redshift, luminous, infrared galaxies 
at submillimeter wavelengths  (Hughes et al. 1998). Indeed the steep rising 
of the X--ray and far--infrared spectrum towards high frequencies 
of the ultraluminous infrared galaxy IRAS 09104+4109 is very similar. 
Conversely, the weak rest--frame optical--UV emission is shifted in the
R band explaining the extremely faint optical magnitudes.  
As a consequence the optical to X--ray flux ratio changes in a non--linear 
way.

We have computed the optical magnitude in the R band
and the 2--10 keV X--ray flux which would 
be observed for a source with the SED of Figure 2 
from $z=0$ to $z=1.5$.
The redshift tracks in the optical magnitude versus X--ray flux plane 
(Fig.~3) have been normalized to the observed X--ray flux and R magnitude
of NGC 6240 and IRAS 09104$+$4109.
The two objects are characterized by a similar SED 
but different X--ray luminosities (about $3 \times 10^{44}$ erg s$^{-1}$ 
and 10$^{46}$ erg s$^{-1}$ respectively).
The results clearly indicate that the observed high values of
the X--ray to optical flux ratio are consistent with those 
expected by a population of high redshift, mildly Compton thick AGN with 
X--ray luminosities in the range log$L_X = 44-46$ erg s$^{-1}$.

We have also tried to explain with a similar approach the 
distribution of $f_X/f_{opt}$ values of {\tt XBONG}
assuming that the underlying SED of the AGN powering the 
hard X--ray emission is that of a heavily Compton--thick object.
The lower redshift track in figure 3, normalized to the locus of 
nearby objects in the BeppoSAX survey of Maiolino et al. (1998),
is in relatively good agreement with the observed $f_x/f_{opt}$
and redshift distribution of the {\tt XBONG} sample.    

Although it seems reasonable to argue that a large fraction 
of {\tt unconventional} AGN are obscured by Compton thick gas 
our approximations appear to be too much simple to draw 
quantitative conclusions.
At the face value our model predicts that 
most of the sources with log$f_X/f_{opt} > $ 1 are 
mildly Compton thick AGN in the redshift range 0.5--1.5.
The presence of a large population of Compton thick 
AGN at faint optical magnitudes has been put forward by 
Fabian et al. (2002). On the basis of a more sophisticated model 
they suggest that such a population would have redshifts 
ranging from 2 to 8 and could be detectable in {\it Chandra} deep fields.

The detection of strong FeK$\alpha$ features
could in principle provide a powerful tool 
to check the space density and redshift distribution 
of Compton thick AGN.
Although only an handful of objects 
exhibit obvious K$\alpha$ lines (Bauer et al this volume) 
we expect to obtain more stringent constraints 
from an undergoing systematic spectral analysis of a 
carefully selected sample of {\tt unconventional} AGN.

\begin{figure*}
\resizebox{\hsize}{!}
{\includegraphics[width=16cm,height=14cm]{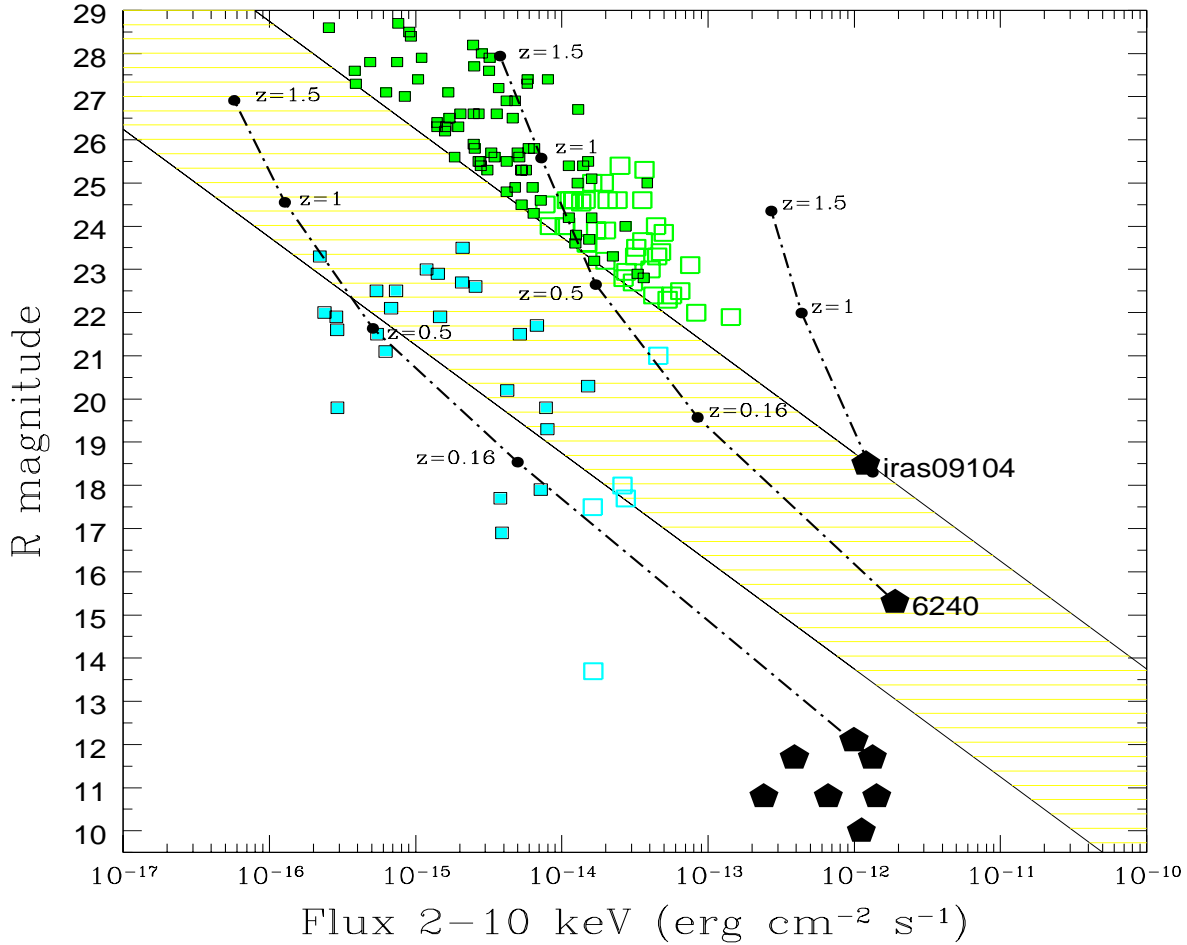}}
\caption{\small The 2--10 keV flux versus the R band magnitude for a sample of 
{\tt unconventional} AGN detected in the {\tt HELLAS2XMM} 
survey (empty squares) and in deeper {\it Chandra} and XMM--{\it Newton} 
surveys (filled squares). 
Green and cyan symbols refer to objects with $f_X/f_{opt} > 1$ 
and {\tt XBONGs}, respectively.
The dash dotted lines represent the redshift 
tracks computed as described in the text. The filled symbols in the lower
part of the diagram correspond to nearby heavily Compton thick AGN
(Maiolino et al. 1998).}
\label{label3}
\end{figure*}

\acknowledgements
 This research has been partially supported by ASI contracts
I/R/113/01; I/R/073/01 and by the MIUR grant Cofin--00--02--36.

\end{document}